\documentclass[a4paper,12pt,english]{article}

\usepackage[pdftex]{graphicx}
\usepackage{amssymb}
\usepackage{multimedia}
\usepackage{amsmath,amsthm}
\usepackage{amssymb}
\usepackage{epstopdf}
\usepackage{times}
\usepackage{mathrsfs}
\usepackage{color}
\usepackage{gensymb}
\usepackage{appendix}
\usepackage{subfigure}

\title{Origin of stabilization of macrotwin boundaries in martensites}

\author{O. U. Salman$^{\rm a,}$\footnote{
Laboratoire  de  M\'ecanique  des Solides,  CNRS-UMR  7649,
Ecole Polytechnique, Route de Saclay, 91128 Palaiseau, France, e-mail: \texttt{umut.salman@polytechnique.edu} }, B. Muite$^{\rm b,}$\footnote{Department of Mathematics,
University of Michigan,
Ann Arbor,
MI 48109-1043, USA, e-mail: \texttt{muite@umich.edu}
} and A. Finel$^{\rm a}$\footnote{ e-mail: \texttt{alphonse.finel@onera.fr}}}
\begin{document}
\maketitle
\noindent\textsuperscript{a} Laboratoire d'Etude des Microstructures, ONERA, BP 72, 92322  Chatillon, France \\
\textsuperscript{b} Mathematical Institute, 24-29 St. Giles, University of Oxford, Oxford, OX1 3LB, UK

 %\ead{umut.salman@polytechnique.edu}
 
 %\author{B. K. Muite}
 %\ead{muite@umich.edu}
 
% \author{A.~Finel}
 %\ead{alphonse.finel@onera.fr}
 
 %\cortext[cor2]{Principal corresponding author}
% \fntext[fn1]{Laboratoire  de  M\'ecanique  des Solides,  CNRS-UMR  7649,
%Ecole Polytechnique, Route de Saclay, 91128 Palaiseau, France}
% \fntext[fn2]{Department of Mathematics,
%University of Michigan,
%Ann Arbor,
%MI 48109-1043}
%
% 
% \address[rvt]{Laboratoire d'Etude des Microstructures, ONERA, BP 72, 92322  Chatillon, France}
% \address[focal]{Mathematical Institute, 24-29 St. Giles, University of Oxford, Oxford, OX1 3LB, UK}

%\author{O. U. Salman$^{\rm a,}$\footnote{Laboratoire  de  M\'ecanique  des Solides,  CNRS-UMR  7649,
%Ecole Polytechnique, Route de Saclay, 91128 Palaiseau, France.}, B. Muite$^{\rm b,}$\footnote{Department of Mathematics
%University of Michigan
%Ann Arbor
%MI 48109-1043
%USA.} and A. Finel$^{\rm a}$$^{\ast}$
%\\\vspace{6pt} } 
%^{\rm a}${\em{Laboratoire d'Etude des Microstructures, ONERA, BP 72, 92322  Chatillon, France}}
%^{\rm b}${\em{Mathematical Institute, 24-29 St. Giles, University of Oxford, Oxford, OX1 3LB, UK}}
%%\\\vspace{6pt}}
%
%
%
%
%\address{$^{\rm a}${\em{Laboratoire d'Etude des Microstructures, ONERA, BP 72, 92322  Chatillon, France}}\\
%$^{\rm b}${\em{Mathematical Institute, 24-29 St. Giles, University of Oxford, Oxford, OX1 3LB, UK}}}

\begin{abstract}
The origin of stabilization of complex microstructures along macrotwin boundaries in martensites is explained by  comparing two models based on Ginzburg-Landau theory. The first model incorporates a geometrically nonlinear strain tensor to ensure that the Landau energy is invariant under rigid body rotations, while the second model uses a linearized strain tensor under the assumption that deformations and rotations are small. We show that the approximation in the second model does not always hold for martensites and that the  experimental observations along macrotwin boundaries can only be reproduced by the geometrically nonlinear (exact) theory.
\end{abstract}

%\begin{keyword}
%martensite \sep patterning \sep twinning \sep phase-field method \sep shape memory alloys 
%\end{keyword}

%\end{frontmatter}

\markboth{U. Salman, B. Muite and A. Finel}{Philosophical Magazine}

\newcommand{\noi} {{\noindent}}

\newcommand {\be}  {\begin{equation}}
\newcommand {\bm}  {\bold}
\newcommand {\ee}  {\end{equation}}
\newcommand {\bea} {\begin{eqnarray}}
\newcommand {\eea} {\end{eqnarray}}

\newcommand {\bit} {\begin{itemize}}
\newcommand {\eit} {\end{itemize}}

%\maketitle

%\begin{abstract}
%Two dynamical models for understanding the role of rigid body rotations in pattern formation in materials undergoing martensitic phase transformation are compared.  Both models are based on a non local continuum elasticity approach for the elastic displacement field describing a square to rectangle transition  using Ginzurg-Landau theory including inertial and Rayleigh dissipation terms. The first model incorporates a geometrically nonlinear strain tensor to assure that the Landau energy is invariant under rigid body rotations, while the second model uses a linearized strain tensor under the assumption that the deformations and rotations are small. The approximation in the second model does not always hold for martensitic phase transformation and may be inappropriate to predict the complex microstructures along macrotwin interfaces observed in experiments, where rotations are quite large. We show that the model with geometrically nonlinear strains produces final metastable states with multiple laminates, separated by stable macrotwin interfaces along which bending, tapering and splitting of microtwin needles is observed, in agreement with experiment. On the other hand, the linear model produces final states that are always simple laminates.
%
%
%\end{abstract}

%\pacs{81.30.Kf,81.40.Jj,62.20.fg,64.70.kd,46.05.+b,46.35.+z}% PACS, the Physics and Astronomy

\section{\label{sec:level1}Introduction}

Martensitic phase transformations are displacive first order transitions that involve a shear dominated change of shape in the underlying crystal lattice from a high symmetry phase, the austenite, to a degenerate low symmetry phase, the martensite. Generally, the martensitic transition leads to very complex self-similar patterns that consist of twinned laminates on a wide range of length scales. It has long been known that an important mechanism at the origin of these complex morphologies is the lattice mismatch between the different variants of the low-symmetry martensite and the austenite. Specifically, lattice continuity at the interfaces between two variants and between any variant and the austenite enforces long-range elastic interactions. The resulting strain energy can relax only through specific arrangements of twinned laminates that lead to a stress-free state characterized, at the scale of the laminates, by an invariant plane strain and, at a larger scale, by a very small average strain (see for example \cite{Bha03} for a general review).

The dynamics of the transition is very often modeled through the use of a Time Dependant Ginzburg Landau theory which typically postulates that strain rates are proportional to a driving force derived from a thermodynamic potential supplemented by compatibility-induced long-range elastic interactions. This overdamped description indeed generates complex microstructures, because it incorporates a self-accomodation mechanism through highly anisotropic long-range elastic interactions. However, when the primary order parameter that controls the transition is an elastic long wavelength mode, as in proper martensites, this overdamped description is inadequate, as it does not incorporate the correct dynamics of the long-wave length sound waves. Indeed, using a simple linear analysis, it is easy to show that, because of the conservation law of density,  the long-wavelength dynamics in the presence of damping leads to
\be \nonumber
\omega(q) \sim v_s q + i\frac{v_s^2}{\lambda} \gamma\, q^2
\ee
where $v_s$ is the sound velocity, $\lambda$ some typical elastic constant and $\gamma$ a damping coefficient. For sufficiently long wavelength and finite sound velocity, the period of oscillation $1/v_s q$ is always smaller than the lifetime $\lambda/v_s^2\gamma q^2$. The corresponding waves are underdamped and, therefore, propagate. In other words, we must consider that the sound velocity is finite and that strain-induced elastic interactions cannot reach their long-range character instantaneously. A correct description of the dynamics of proper martensites must therefore incorporate inertial effects, which requires the use of an inertial Lagrangian description.

Another important feature of martensites is the presence of lattice rotations. More precisely, the transformation strain from austenite to martensite may be very large (up to the order of 10\%) and thus lattice rotations may be important. This requires the use of a model that does not penalize rotations and  hence is  rotationally invariant. A model is considered to be invariant under rotations if its energy and dissipation potential are unchanged when its deformation field is rotated (see for example \cite{Ant05,Dem00} and the references therein). Most dynamic simulations of the martensitic transition have used strain-based models that are geometrically linear, i.e. that assume a linear relationship between the displacement and strain fields. This linearization simplifies greatly the  formalism and allows the use of simple and stable numerical schemes. However, linear geometry is obtained as an approximation of the exact geometry  under the assumption that displacements are infinitesimally small. As a consequence, the description is not invariant under rigid body rotations, in contrast to nonlinear geometry, which preserves the exact nonlinear relationship between the displacement and strain fields and is therefore rotationally invariant. 

Analytical studies of static energy minimizing geometrically nonlinear models and their geometrically linear counterparts indicate that there are significant differences between the two models and in idealized conditions the latter can even reproduce qualitatively incorrect results ~\cite{Bha03,DolMul95a,DolMul95c}. However, the effects of geometrical nonlinearities on the dynamics in realistic conditions are still an open question. One may easily argue that neglecting these nonlinearities may have drastic consequences in an iterative process.

 In this paper,  we present a numerical comparison of geometrically nonlinear and linear modeling for identical physical problems from thermodynamical, elastic and dissipative point of view and for the same physical parameters (mass density, elastic constants, damping coefficient and interfacial energy).  We show that dynamic models for pattern formation in martensites with geometrically nonlinear strains capture physically relevant features not captured in geometrically linear models. In particular, we show that the experimentally observed macrotwin interfaces can only be reproduced within a nonlinear theory, as long-lived metastable states which cannot be reached in the geometrically linear model.

\section{\label{sec:level1}Viscoelastic models for martensitic phase transformations}
A martensitic transformation is characterized by a displacement field $u(x)=y(x)-x$, where $y(x)$ and $x$ are the coordinates of a material point in the deformed and undeformed states, respectively. Elastic energy $F_L$ depends only on the deformation gradient $\bm{F}= \bm I + \nabla{u}$ and should be rotationally invariant. A simple way to enforce this property is to assume that $F_L$ is only function of the Lagrangian tensor $\bm{\epsilon^{NL}}= \left(\bm{F}^T\bm{F}-\bm{I}\right)/2$, which amounts to assuming, as usual, that the elastic energy depends only on the length changes between material points.  This functional dependance guaranties that $F_L$ is invariant under rigid body rotations, simply because the strain tensor $\bm \epsilon^{NL}$ itself is invariant if a rotation $\bm Q$ is applied to the deformed state:

\be
\bm{\tilde \epsilon}^{NL}=\frac12 \left(\bm{(\bm QF})^T(Q\bm{F)}-\bm{I}\right)=  \frac12 \left(\bm{F}^T\bm{F}-\bm{I}\right)= \bm \epsilon ^{NL}.
\ee
We use the superscript $NL$ to stress the fact that the Lagrangian strain is a nonlinear  function of the displacement gradient :

\be
\label{nonlin}\bm{\epsilon}^{NL}_{ij}= \frac{1}{2}
\biggl(\frac{\partial u_i}{\partial x_j}
+
\frac{\partial u_j}{\partial x_i}
+
\frac{\partial u_k}{\partial x_i}
\frac{\partial u_k}{\partial x_j}
\biggr).
\ee
The linear geometry approximation consists in neglecting the quadratic terms in the previous equation \cite{Bha03,LanLif84}, leading to a linear strain tensor $\bm \epsilon ^L$:
\be
\label{lin}\bm{\epsilon}^{L}_{ij}= \frac{1}{2}
\biggl(\frac{\partial u_i}{\partial x_j}
+
\frac{\partial u_j}{\partial x_i}
\biggr)
\ee
i.e. $\bm{\epsilon^{L}}= \frac12\left(\bm{F}+\bm{F}^T\right)-\bm{I}$. Obviously, $\bm \epsilon^L$ is not rotationally invariant.

 In this paper, we consider the square to rectangle transformation as the simplest two dimensional model case for pattern formation in martensites (we will comment on that point below). The symmetry adapted linear combination of strain components are:
\begin{align}
&{} e_1=\epsilon_{11}+ \epsilon_{22}, \quad e_2= \epsilon_{11}-\epsilon_{22}, \quad e_3= \epsilon_{12}= \epsilon_{21}.
\label{order_parameters}
\end{align}
The quantity $e_2$ corresponds to the deviatoric strain and is the primary order parameter, as it controls the square to rectangle transformation. The first and third terms $e_1$ and $e_3$ correspond to the dilatational and shear strains, respectively, and play the role of secondary order parameters. To the lowest order, the corresponding  strain energy density is given by the following Landau potential:

\begin{align}F_L=a_2e^2_{2} + a_4e^4 _{2}+a_6e^6 _{2}  +b_1e^2_{1}+b_3e^2_{3}. 
\label{strain_energy}
\end{align}

\noi The coefficients $a_2$, $a_4$, $a_6$, $b_1$, and $b_3$ can be tuned to reproduce the elastic constants and the stress-free transformation strain of a specific alloy. We also include a gradient contribution, in the form of a Ginzburg term, in order to penalize interfaces between variants. This term of course breaks the scale invariance of the elastic medium and provides us with a length scale. For the sake of simplicity, we used here the following simple nonlocal density:
\begin{align}
  F_G = \frac{\beta}{2}\left(\Delta u_i\right)^2 \label{Ginzburg}
\end{align}
 which fulfills the constraint of being rotationally invariant. Finally, we include a dissipation mechanism through a Rayleigh dissipation density given by:
\be
R = \frac12 \gamma_i \dot e_i^2
\label{Rayleigh}
\ee
where, for simplicity, the viscosity coefficients $\{\gamma_i\}$ are supposed to be independent of the amplitudes of the local order parameters $e_i$. 
Taking into account a kinetic energy density $T = \frac12 \rho \dot u_i$, we construct the Lagrangian density $L = T-F_L -F_G$. 
Finally, the dynamics of the system is given by  the Langrange-Rayleigh equations:
\be
\frac{d}{dt} \frac{\partial L}{\partial \dot u_i} - \frac{\partial L}{\partial  u_i} = - \frac{\partial R}{\partial  \dot u_i}.
\ee
%
%\noi  where $\mathcal L=\int L\, d^2x$ and $\mathcal R = \int R \,d^2x$ are the Lagrangian and dissipation function of the system, respectively. 
%
These dynamical equations may be written in the general form:
\be
\rho \,\ddot u_i = \frac{\partial}{\partial x_j} \frac{\partial F_L}{\partial u_{i,j}} - \frac{\partial}{\partial x_j}\frac{\partial}{\partial x_k} \frac{\partial F_G}{\partial u_{i,jk}} +\frac{\partial}{\partial x_j}\frac{\partial R}{\partial \dot u_{i,j}} 
\ee
where as usual $u_{i,j}$ and $u_{i,jk}$ stand for $\frac{\partial u_i}{\partial x_j}$ and  $\frac{\partial^2 u_i}{\partial x_j\,\partial x_k}$, respectively.

This formulation fits both for geometrically nonlinear and linear models. These two models differ only in the way the strain components $\epsilon_{ij}$ enter, through the order parameters $e_i$, into the definitions of the local strain energy $F_L$ and Rayleigh dissipation term $R$. They are linked to the displacement field $u_i$ by Eq. \ref{nonlin} for the nonlinear model and Eq. \ref{lin} for the linear one.

We have chosen these equations as the simplest geometrically nonlinear models, which can be compared to recent geometrically linear implementations of phase field (without inertia)~\cite{KerEtAl99, ArtJinKha01,Wang:2004lf} and Lagrangian dynamics~\cite{CunJac01,LooEtAl03,AhlLooSax06} models for martensitic phase transformations.
In the following simulations, the dissipation term defined in Eq. \ref{Rayleigh} is treated linearly (i.e. with Eq. \ref{lin}) for numerical simplicity. The effect of a nonlinear dissipation term is discussed elsewhere~\cite{Sal08,Muite:2009kl}. However, we mention here that the models with linear and nonlinear dissipation terms lead to very similar final microstructures, simply because dissipation is vanishingly small in the late stage of the dynamics.
\section{\label{sec:level2}Experimental motivation}

\begin{figure}
\begin{center}
\subfigure[$\bold{t=20\times10^4}$][]{\label{macrotwin_in_NiAl:edge-a}\includegraphics[height=5cm]{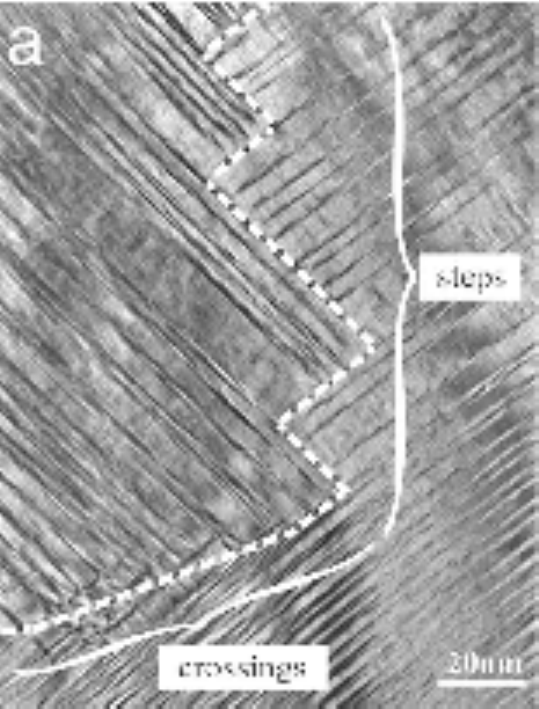}}
\subfigure[$\bold{t=30\times10^4}$][]{\label{macrotwin_in_NiAl:edge-b}\includegraphics[height=5cm]{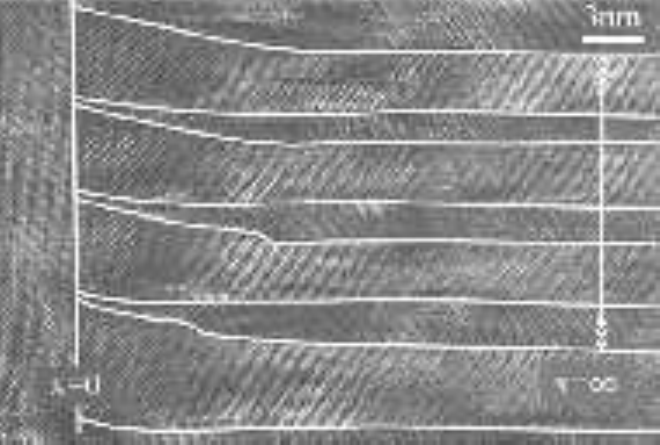}}
\end{center}
\caption{\label{macrotwin_in_NiAl}(a) Typical TEM image of a macrotwin boundary in Ni$_{65}$Al$_{35}$ revealing step type microstructures. (b) Typical HRTEM image of nanoscale martensite needles close to one of the steps of the macrotwin image shown in (a). Taken from  \cite{Boullay:2001rw}. }
\end{figure}

We want here to investigate the dynamics and morphology of macrotwins, such as those observed in Ni$_{65}$Al$_{35}$ \cite{SchEtAl02,Boullay:2001rw,SchBouKohBal01}, a shape-memory alloy which undergoes a cubic to tetragonal martensitic transition upon cooling (see Fig. \ref{macrotwin_in_NiAl}). These configurations are very common in martensites, as they are one of the building blocks of the elastic energy minimizing microstructures generally observed in the late stage of the transformation. We will here specify the discussion to the cubic to tetragonal transformation, because this situation may be compared to the 2D square to rectangle transition modeled in this paper (see the discussion below). 

\begin{figure}
\begin{center}
\subfigure[$\bold{t=20\times10^4}$][]{\label{twin_walls:edge-a}\includegraphics[height=6cm]{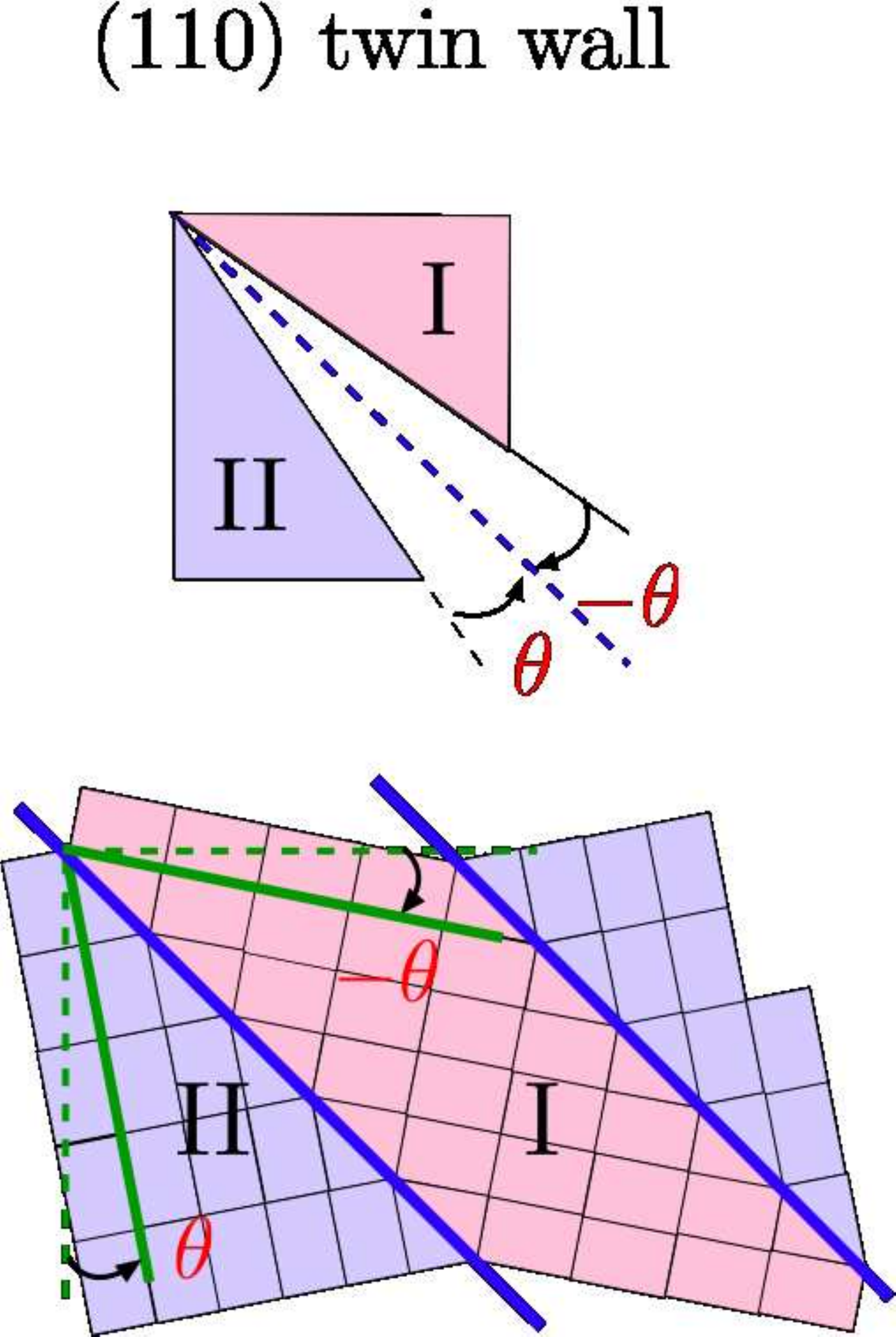}} \hspace{2cm}
\subfigure[$\bold{t=30\times10^4}$][]{\label{twin_walls:edge-b}\includegraphics[height= 6cm]{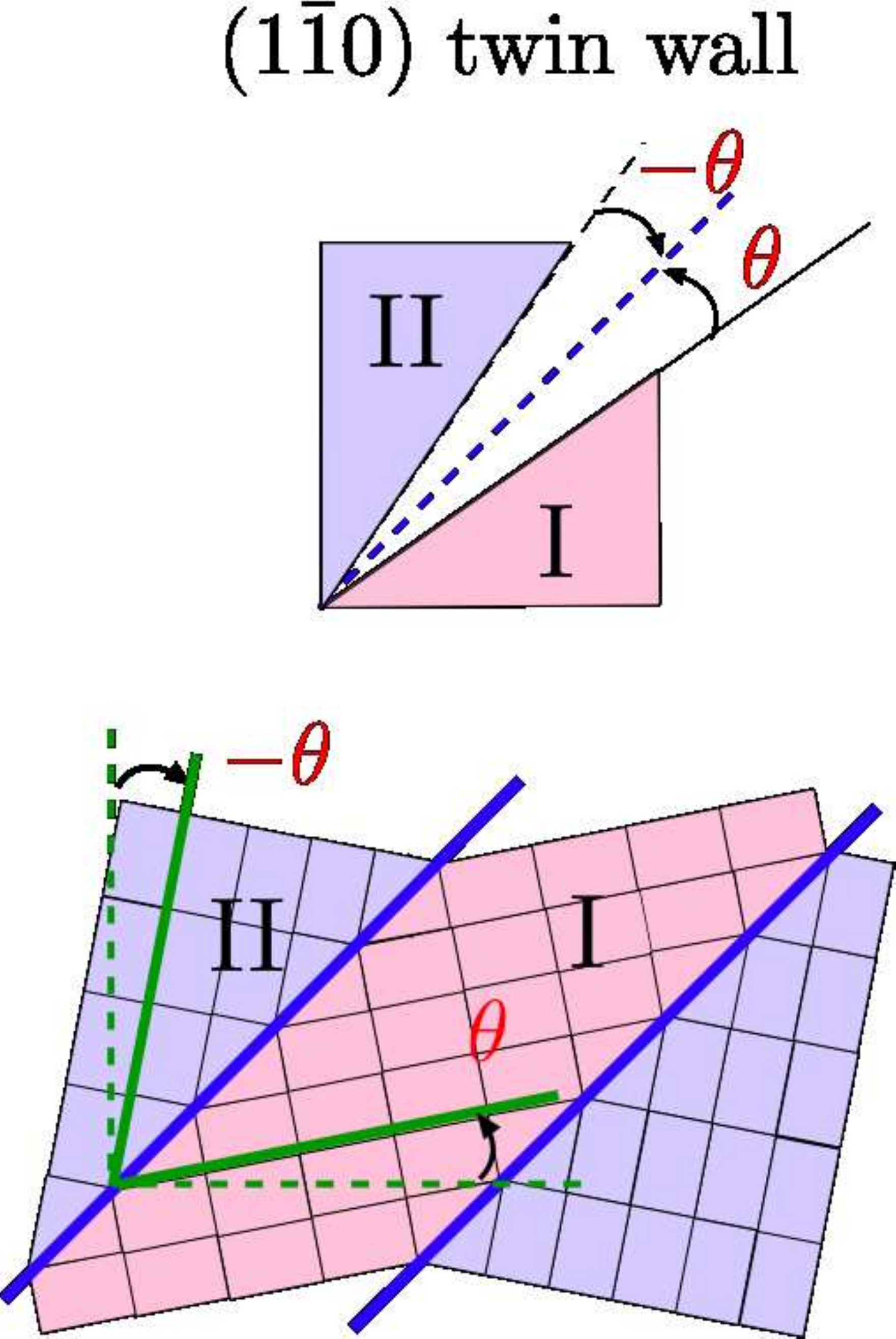}}
\end{center}
\caption{\label{twin_walls} Twin walls involving variant I and II, whose tetragonal axis lie along directions [100] and [010], respectively. (a): the twin wall is parallel to a (110) plane of the austenite. (b): the twin wall is parallel to a $(1 \bar 1 0)$ plane. In the case of Ni$_{65}$Al$_{35}$, we have $\theta \sim 6 \degree$. }
\end{figure}

Generally speaking, when the martensitic transition proceeds, multiply twinned martensitic plates are formed to accomodate the shape change and minimize the elastic energy. Each plate consists of two of the three tetragonal L1$_0$  variants, separated by microtwin walls. The microtwin walls inside a plate originate from $\{110\}$-type planes of the B2 austenite. When the transformation proceeds, each plate continues to grow: a macrotwin plane forms whenever two such polytwinned plates come into contact and try to accommodate. We consider here the situation depicted schematically in Fig. \ref{twin_walls}, where two polytwinned plates are formed with alternating sequences of variants I and II, whose tetragonal axis are originally parallel to directions $[100]$ and $[010]$ of the austenite, respectively. A first plate is formed when variants I and II develop alternatively along microtwin walls that originate from former (110) planes of the austenite. As seen in Fig. \ref{twin_walls:edge-a}, this accommodation mechanism requires rotations of variants I and II opposite in sign\footnote{\label{rere}During the growth process, compatibility of the plate with the austenite requires that there is a small rigid-body rotation of the entire plate that destroys this balance.} and which are not small if the shape change is large. More precisely, if  the transformation strain $U$ that transforms the austenite to variant I is written as $^{\ref{rere}}$,
\be
U=\text{Diag}[\beta,\alpha,\alpha]
\label{U_matrix}
\ee
the angle $\theta$ is given by $\tan \theta = \frac{\vert \alpha - \beta \vert }{\alpha+\beta}$. In the case of Ni$_{65}$Al$_{35}$ \cite{Boullay:2001rw,SchBouKohBal01}, the entries of the transformation strain have been  estimated to $\alpha \simeq 0.93$ and $\beta \simeq 1.15$, which leads to $\theta \sim 6 \degree$.  Another possibility for the same variants I and II is to accommodate along  twin walls originating from former ($1\bar{1}0)$ planes of the austenite, as in Fig. \ref{twin_walls:edge-b}. This second configuration requires rotations opposite in sign to the ones involved in the first plate. 

Now, when these two microtwinned plates come into contact along a common interface, this interface often reveals a zig-zag configuration consisting of successive well defined steps, as seen in Fig. \ref{macrotwin_in_NiAl:edge-a}. Along one of these steps, the microtwins of one plate end at a twin wall of the other plate (see Fig.\ref{macrotwin_in_NiAl:edge-b}), this wall thus providing the macrotwin interface, which is locally a prior $\{110\}$ plane of the austenite. At a larger scale, the steps alternate and we observe a zig-zag configuration of macrotwin interfaces parallel  to prior $(110)$ and $(1\bar 1 0)$ planes. An important feature of the macrotwin interface is the specific shape adopted by the twins that are perpendicular to the boundary: as seen in Fig. \ref{macrotwin_in_NiAl:edge-b}, alternate twins narrow and bend when approaching the macrotwin interface.

\section{\label{sec:level2}Numerical Results}
Our aim is to analyze macrotwins associated to the cubic-to-tetragonal transformation, but here we consider the simpler 2D square to rectangle transition. The latter may be interpreted as a 3D cubic to rectangle transition with the following restrictions. Consider a 3D crystal where the displacement field fulfills the following constraints: components $u_1$ and $u_2$ are translationally invariant along axis $x_3$ and the third component $u_3$ varies linearly with $x_3$, i.e. $u_1(x_1,x_2,x_3)=u_1(x_1,x_2)$, $u_2(x_1,x_2,x_3)=u_2(x_1,x_2)$ and $u_3(x_1,x_2,x_3)=\lambda x_3$.  The strain tensor $\epsilon_{ij}$ then adopt the following restricted form:
\begin{equation}
\epsilon_{ij}^{NL}=\left(
\begin{matrix}
\epsilon_{11}(x,y)   & \epsilon_{12}(x,y) & 0 \\
\epsilon_{21}(x,y)   & \epsilon_{22}(x,y) & 0 \\
0 & 0  & \lambda+\frac{\lambda^2}{2}
\end{matrix}\right)
\end{equation} 
where we used the nonlinear formulation. Obviously, if $\lambda = \alpha-1$, where $\alpha$ is defined in Eq. \ref{U_matrix}, any laminate composed of tetragonal variants I and II, and hence with $\{110\}$-type twin walls,  will be stress free. Therefore, from the microstructural and energetic point of views, this laminate is exactly equivalent to a 2D laminate consisting of the two rectangular variants of the square-to-rectangle transition with $\{11\}$-type twin walls. In other words, we conclude that our 2D model should reproduce correctly the morphology of 3D multiple laminate structures formed by tetragonal variants I and II and the corresponding macrotwins, such as those observed in the late stage of the martensitic transition. Of course, we cannot expect that a 2D model will capture correctly the dynamics itself, in particular when austenite and martensite coexist. However, the results reported below concern specifically the stability of macrotwins in the late stage of the transition.

The evolution equations were solved using a Fast Fourier Transform   method with implicit-explicit finite difference timestepping schemes \cite{Sal08,Mui08b,LeVeque:2007zi}. Equations are first written in a dimensionless form using adapted units for mass density ($\rho_0= 6.657$ g/cm$^3$), time ($t_0=10^{-13}$ s), spatial coordinate ($d_0=0.1$  nm) and energy density ($f_0=29.8$ GPa). The dimensionless parameters used here are the following: $\tilde \rho=1$, $\tilde a_2 = -0.40, \tilde a_4 = -2.18, \tilde a_6 = 119, \tilde b_1 = 2, \tilde b_3= 2, \tilde \gamma = 0.1, \tilde \beta = 0.1$, where $\tilde \alpha$ refers to the dimensionless counterpart of the physical quantity $\alpha$. These parameters have been chosen to make our 2D system representative of the situation in Ni$_{65}$Al$_{35}$. This requires the knowledge of experimental quantities, such as the twin interfacial energy and, as we are concerned with microstructures in the martensitic state, elastic constants of the martensite. Unfortunately, as far as we know, these constants are not known.  However, experimental measurements of elastic constants in the austenite are available. In particular, using ultrasonic technics \cite{Davenport:1999gg}, it has been observed that the elastic constants $C_L = (C_{11}+C_{12}+2C_{44})/2$ and $C_{44}$ varies only slightly when approaching the martensitic  transition, contrary to the shear elastic constant $C' = (C_{11}-C_{12})/2$. We thus assumed that the measurements for $C_L$ and $C_{44}$, with $C_L \simeq 290 $ GPa and $C_{44} = 120 $ GPa,  can be used to estimate the corresponding elastic constants in the martensitic phase\footnote{It is understood here that the elastic constants of the martensite are expressed in the undeformed reference state, i.e. in the coordinate axis of the austenite. With the simple strain energy used here (Eq. \ref{strain_energy}), it is easy to show that, within this coordinate system, the elastic constants of the martensite verify $C_{11}=C_{22}$. An inequality would be obtained if a coupling between the dilatational $e_1$ and deviatoric $e_2$ strains was included in the strain energy density.}.  In addition, as we consider here a situation which is far below the transition temperature, the shear elastic constant $C'$ should not be particularly small. We assume that it is of the same order as $C_{44}$. Specifically, we used $C'\simeq C_{44}$ (similar results have been obtained with  $C'\simeq 2C_{44}$). An estimation of the twin interfacial energy is also needed, in order to fix the length scale of the simulation. We chose here $\sigma \simeq 75$ mJm$^{-2}$. We mention that a specific estimate of $\sigma$ is needed only to fix the length scale of the system. Any choice leads to the same dimensionless equations and, therefore, to the same microstructural evolution. Finally, we mention that the choice presented above leads to an equilibrium stress free deviatoric strain close to the experimental one, namely $e_2^{eq}=0.2$.

The simulations are carried out on a grid of size $1024\times1024$ that corresponds to a physical size of $0.1\mu$m$\times0.1\mu$m. The interfaces within the simulation spread typically over three grid points. This corresponds to an interface width of the order of  $0.3$ nm, which is close to typical interface widths observed in the experiments.           

\begin{figure}[htbp]
\begin{center}
\includegraphics[height= 4.5cm]{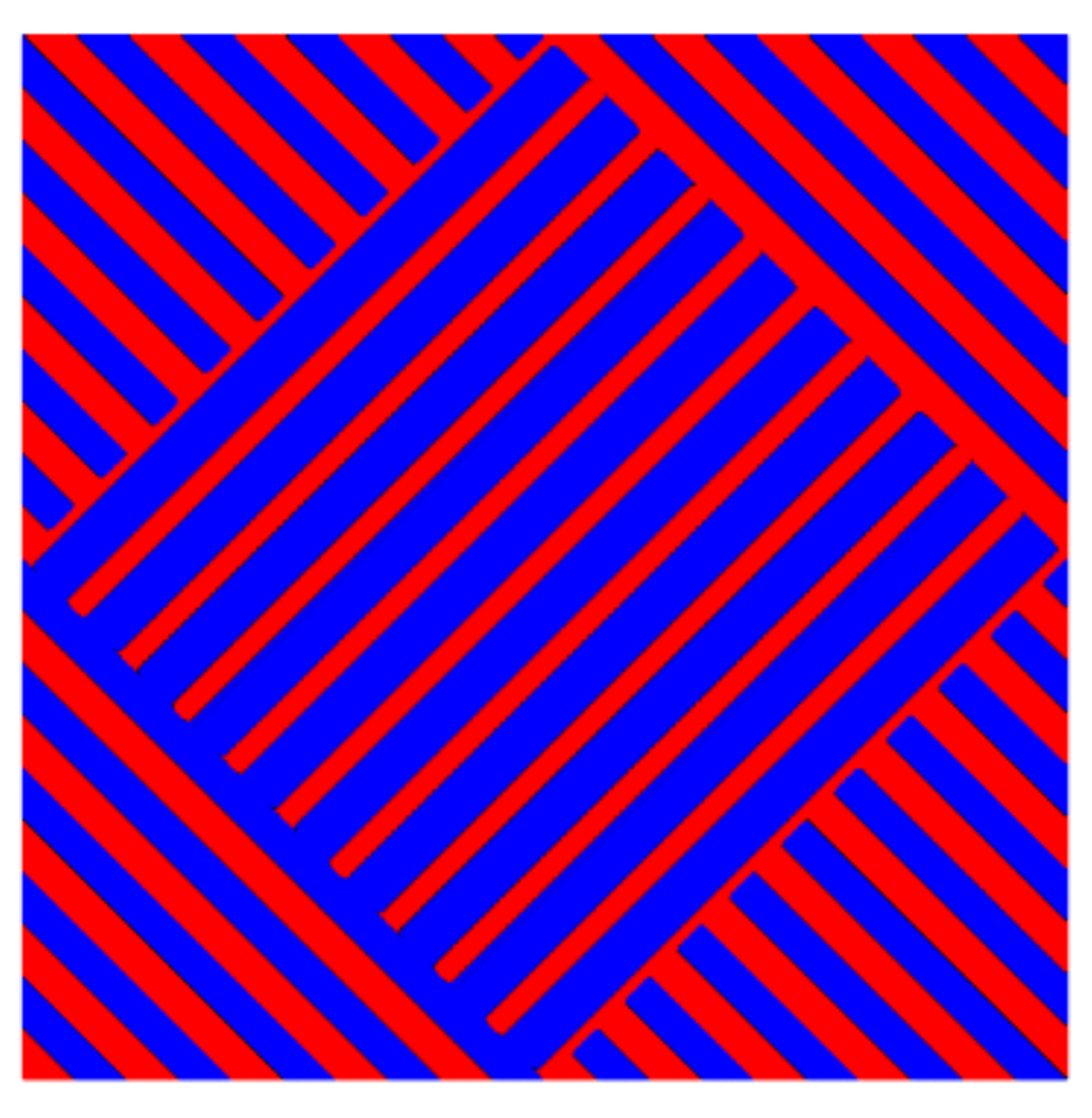}
\end{center}
\caption{\label{initial_configuration}Initial condition for the deviatoric strain $e_2$. The order parameter $e_2$ is initialized in the form of sinusoidal waves of amplitude $0.01$ (the equilibrium deviatoric strain is $e_2 = \pm 0.2$) and $e_1$ and $e_3$ are set identically to zero. Red and blue domains correspond to variant I and II, respectively.}
\end{figure}

Motivated by the experimental situation discussed above, which reveals structural features corresponding to large crystal lattice rotations along macrotwin boundaries \cite{Boullay:2001rw}, the simulations start with the initial configuration shown in Fig. \ref{initial_configuration}. It consists of four macrotwins between laminates formed by alternating sequences of variant I and II. The deviatoric order parameter field $e_2$ is initialized to small values, proportional to the equilibrium value of the corresponding variant, and $e_1$ and $e_3$ are set identically to zero. The initial velocity of the displacement fields $u_i$ is also set to zero.

\begin{figure}
\begin{center}
\subfigure[$\bold{t=20\times10^4}$][]{\label{time_evnl:edge-a}\includegraphics[height=4.5cm]{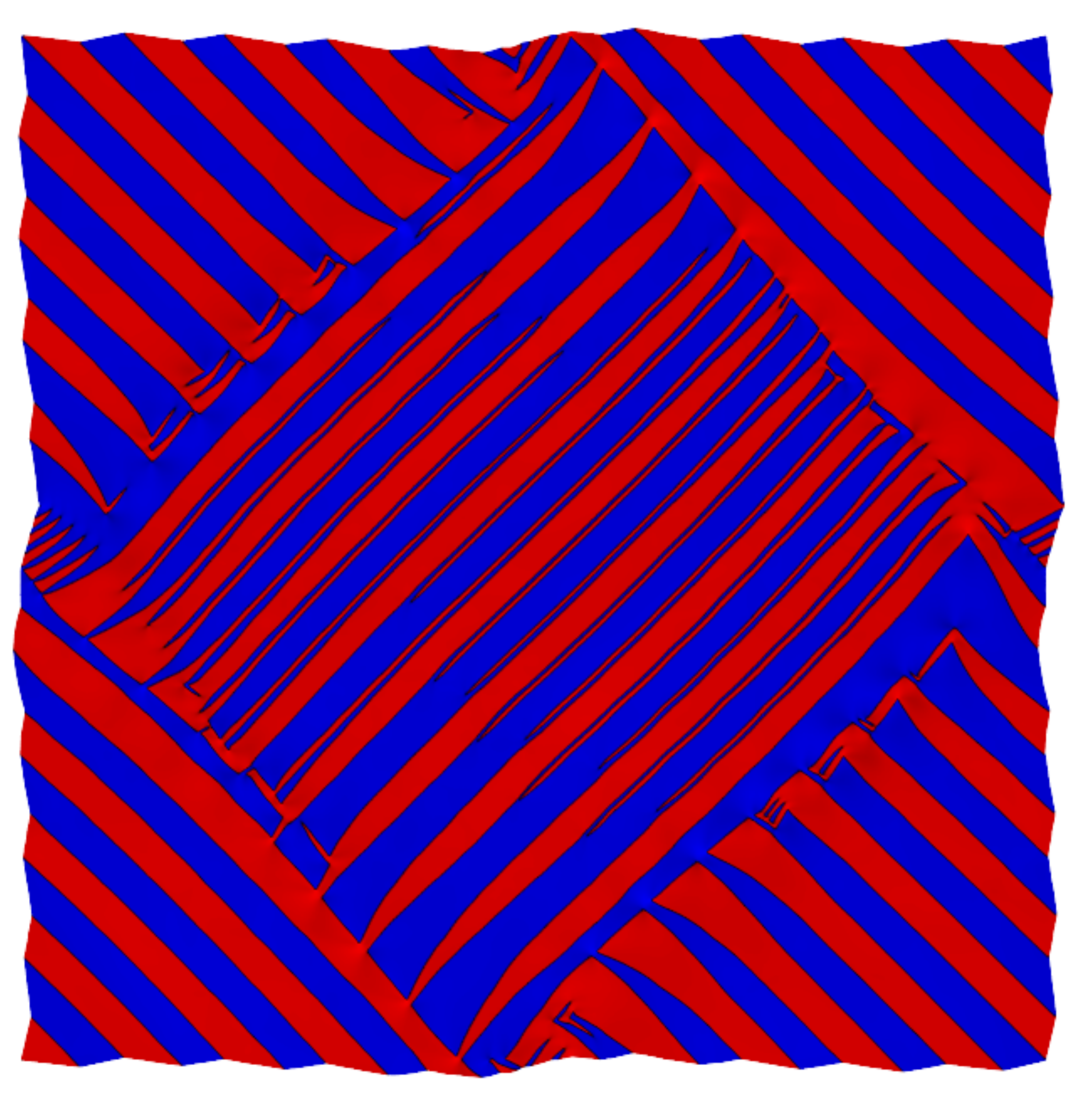}}
\subfigure[$\bold{t=30\times10^4}$][]{\label{time_evnl:edge-b}\includegraphics[height= 4.5cm]{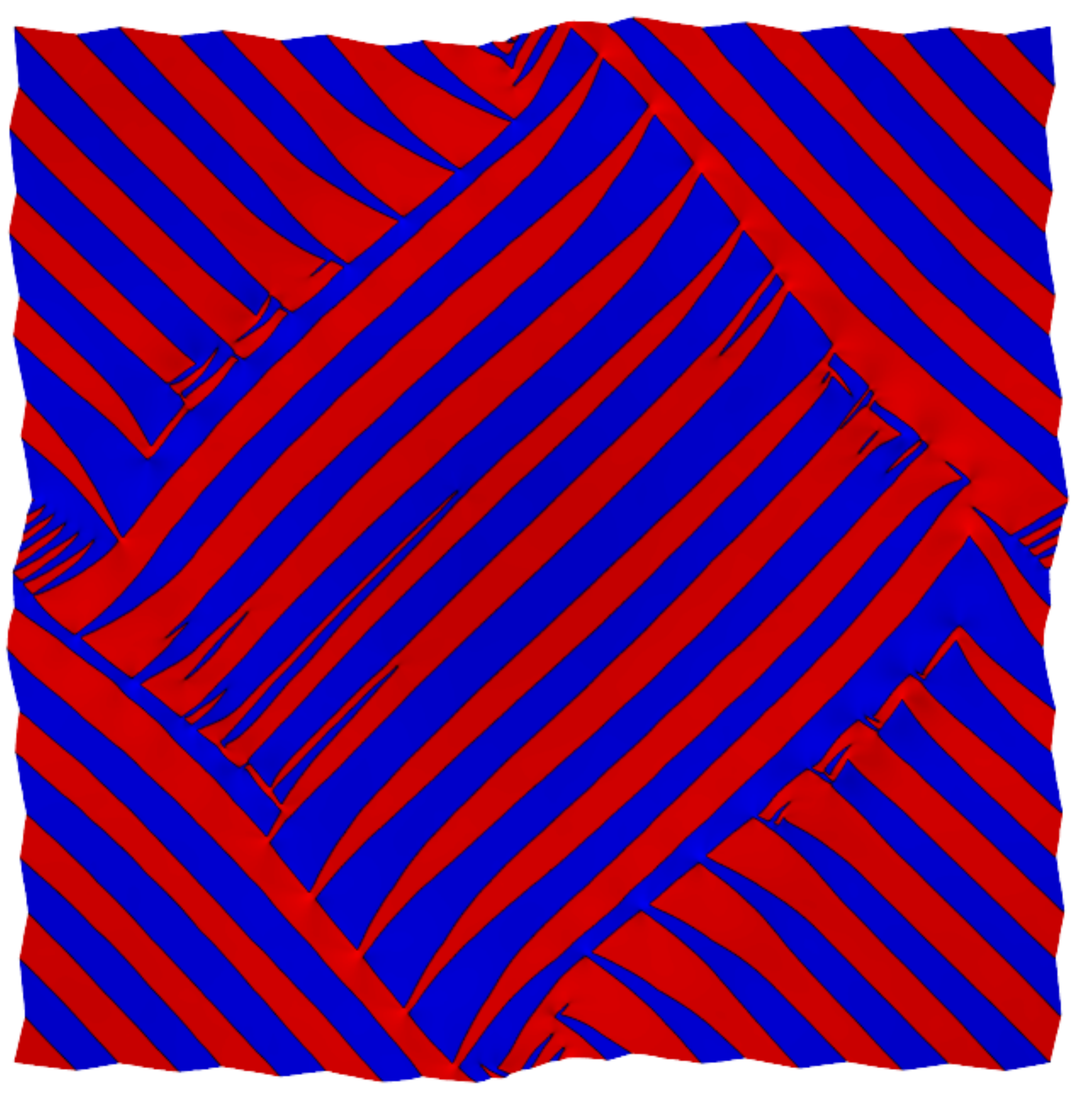}}
\subfigure[$\bold{t=200\times10^4}$][]{\label{time_evnl:edge-c}\includegraphics[height= 4.5cm]{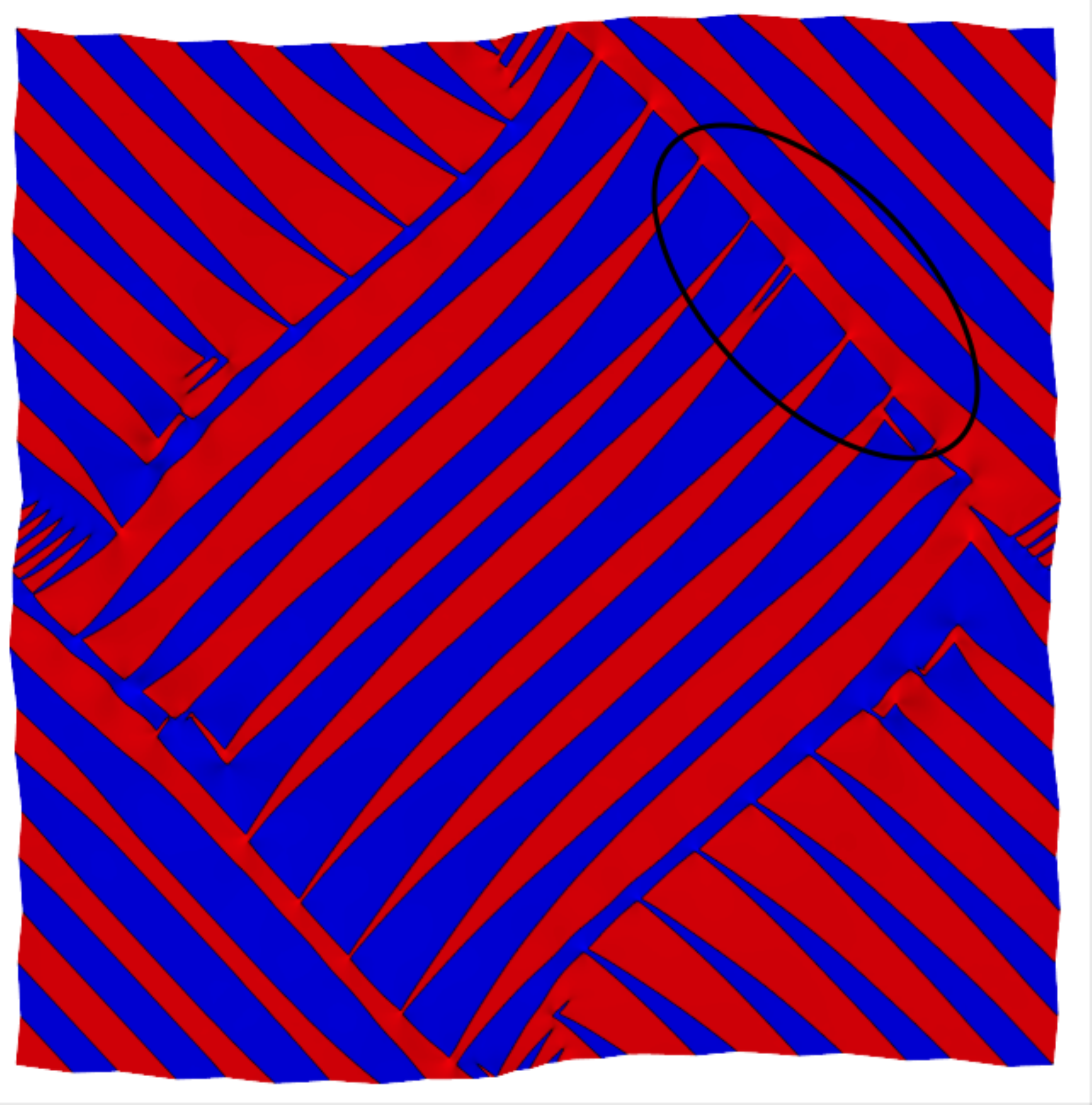}}
\end{center}
\caption{\label{time_evnl}Microstructure evolution in the geometrically nonlinear model. The reduced times from (a)-(c) are 
$\tilde t= 40, 100, 400 \times 10^{2}$.}
\end{figure}

The dynamics is first run within the exact nonlinear geometry scheme. The corresponding microstructural evolution is shown Fig.\ref{time_evnl}. We observe that the microtwins in the neighborhood of macrotwin boundaries display new features at early stages. In particular, a variant that meets a perpendicular lamella made of the same orientational variant is seen to bend and to narrow when approaching the macrotwin interface (see Fig. \ref{time_evnl:edge-a}). Finally, as seen in Fig. \ref{time_evnl:edge-c}, the final state obtained using the nonlinear modeling still displays the initial macrotwin boundaries, but characterized now, on one side, by alternating sequences of variants that narrow and bend when approaching the macrotwin boundary. We mention that this final state, although metastable, does not evolve anymore.

\begin{figure}
\begin{center}
\subfigure[$\bold{t=20\times10^4}$][]{\label{time_evl:edge-a}\includegraphics[height= 4.5cm]{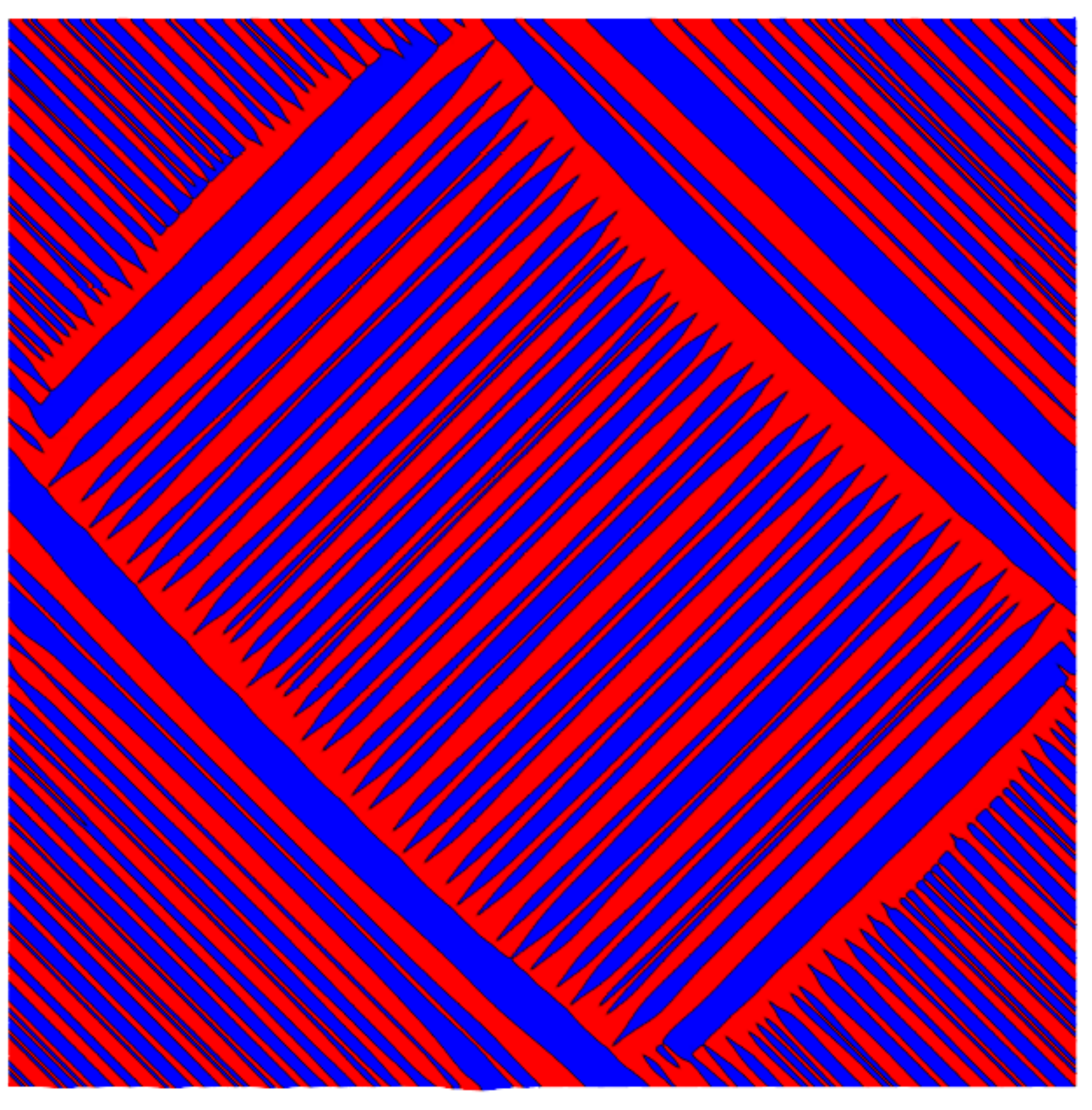}}
\subfigure[$\bold{t=30\times10^4}$][]{\label{time_evl:edge-b}\includegraphics[height= 4.5cm]{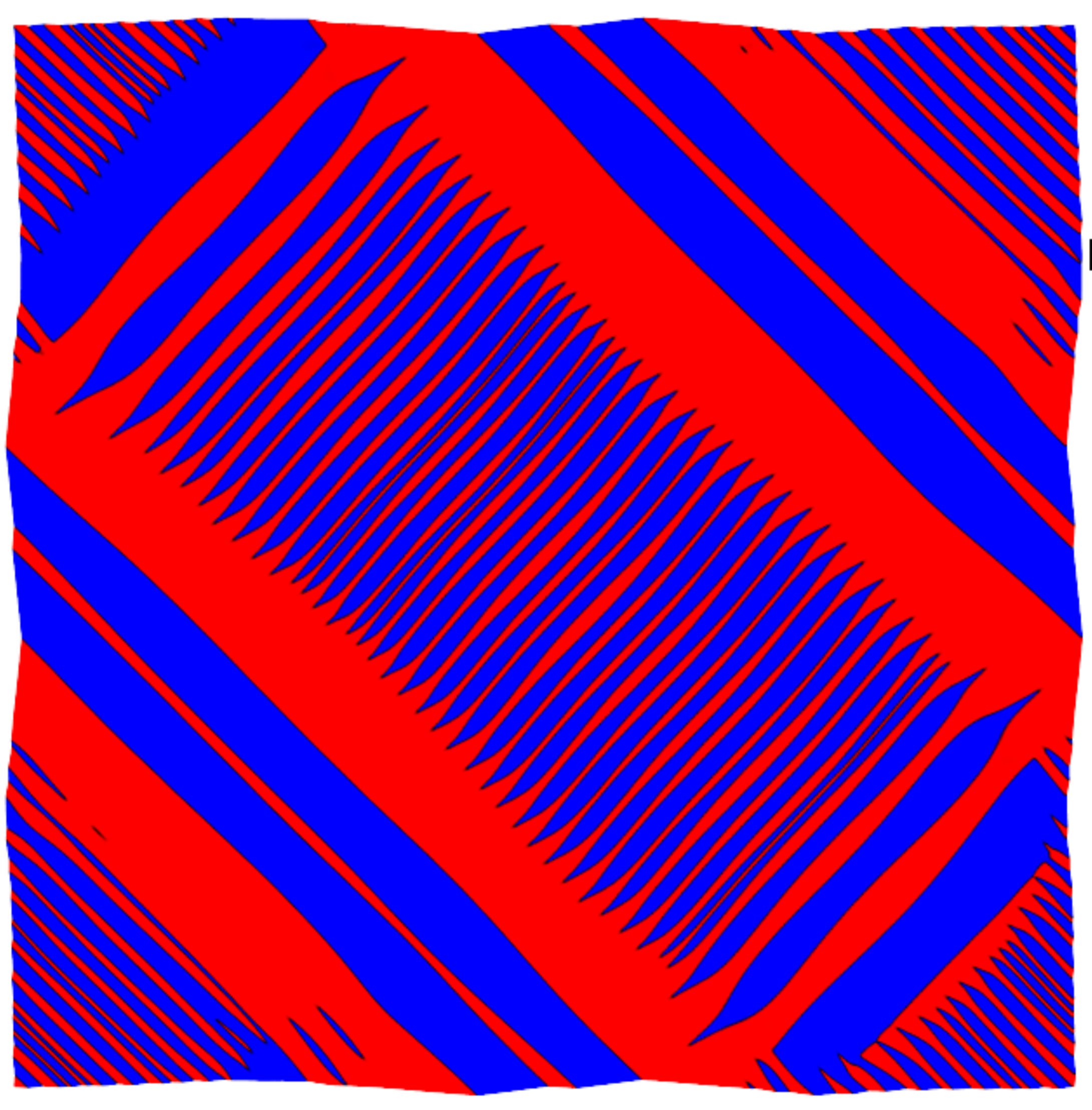}}
\subfigure[$\bold{t=200\times10^4}$  ][]{\label{time_evl:edge-c}\includegraphics[height= 4.5cm]{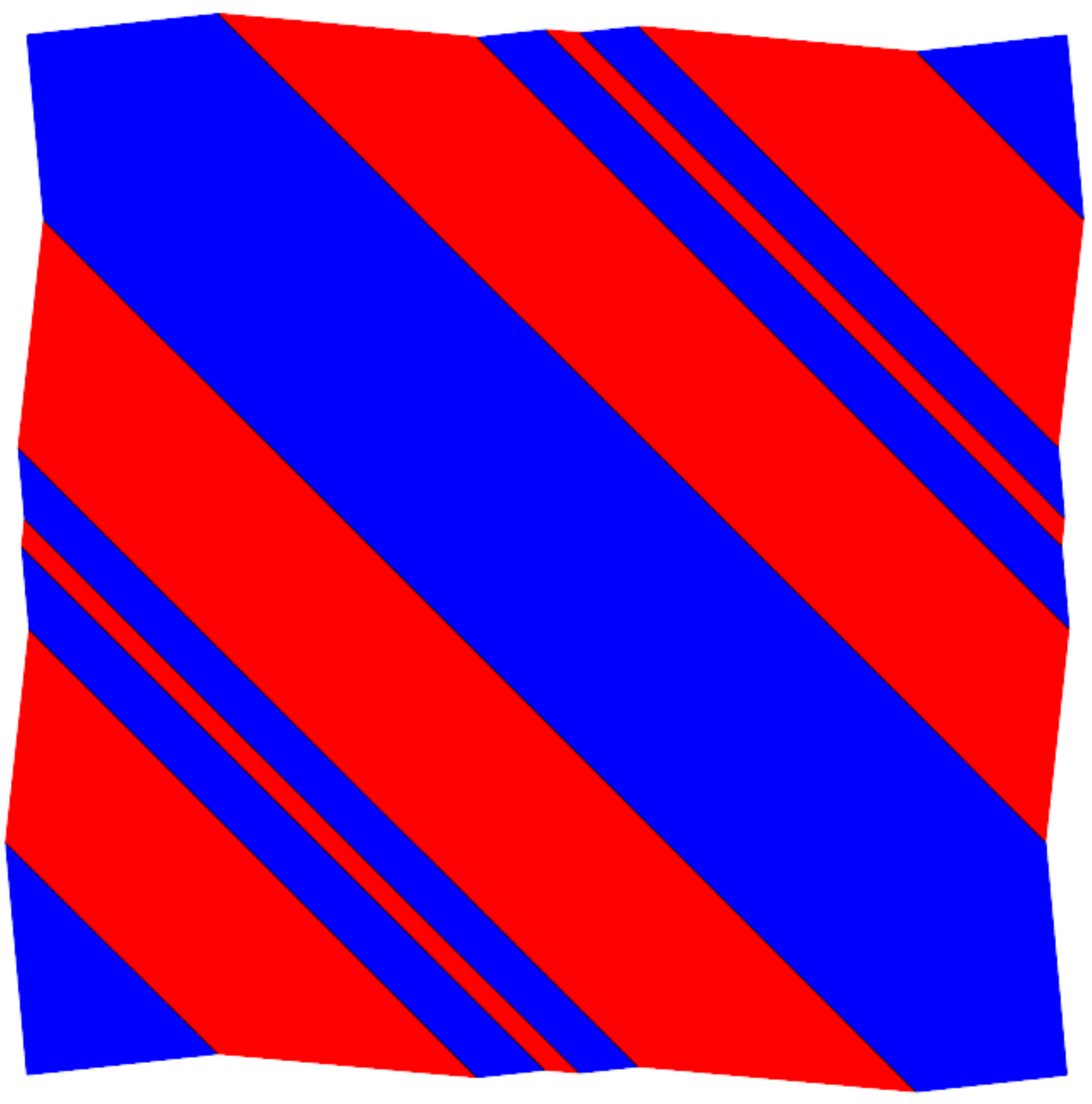}}
\end{center}
\caption{\label{time_evl}Microstructure evolution in geometrically linear model. The reduced times from (a)-(c) are 
$\tilde t= 40, 100, 400 \times 10^{2}$.}
\end{figure}

For comparison, we present in Fig. \ref{time_evl} the microstructural evolution within the linear geometry. We used the same material parameters as for the nonlinear modeling. The only difference is that we used here Eq. \ref{lin} instead of Eq. \ref{nonlin}. We note that the final microstructure is completely different from the one obtained with the linear model. We will comment further on this point below.

\begin{figure}
\begin{center}
\vspace{0.5cm}
\subfigure[][]{\label{simuvsexp:edge-a}\includegraphics[height= 5cm]{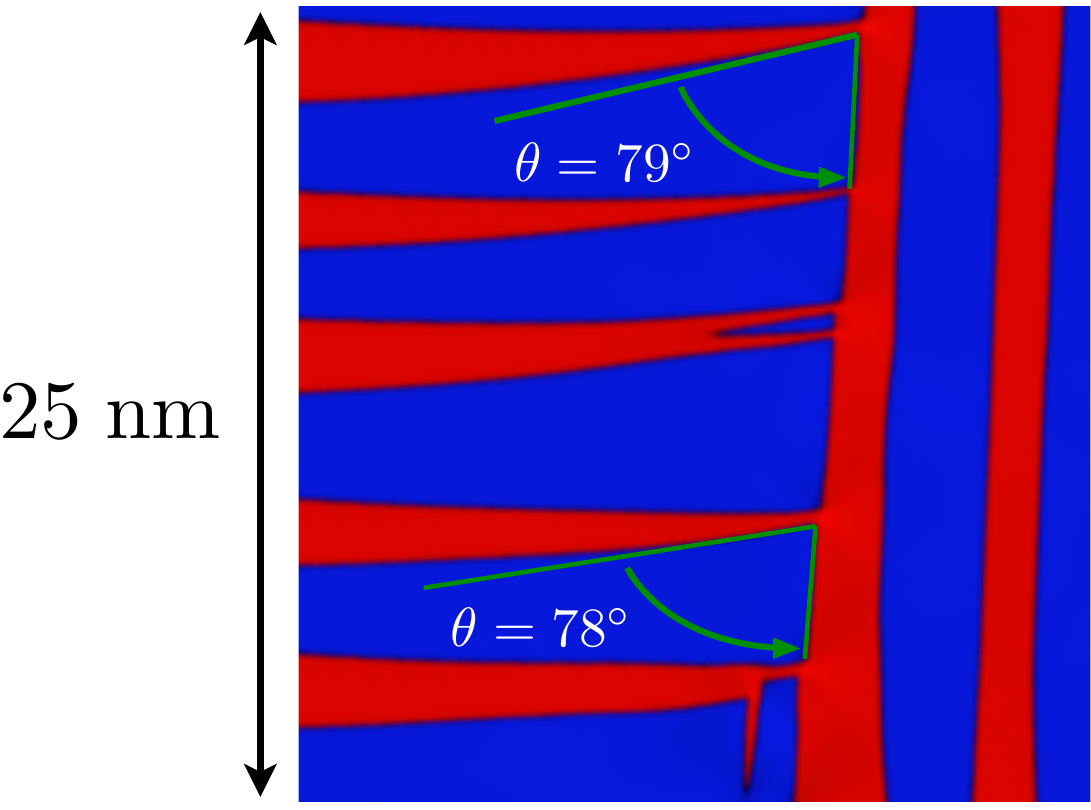}}
\hspace{0.3cm}
\subfigure[][]{\label{simuvsexp:edge-b}\includegraphics[height= 5cm]{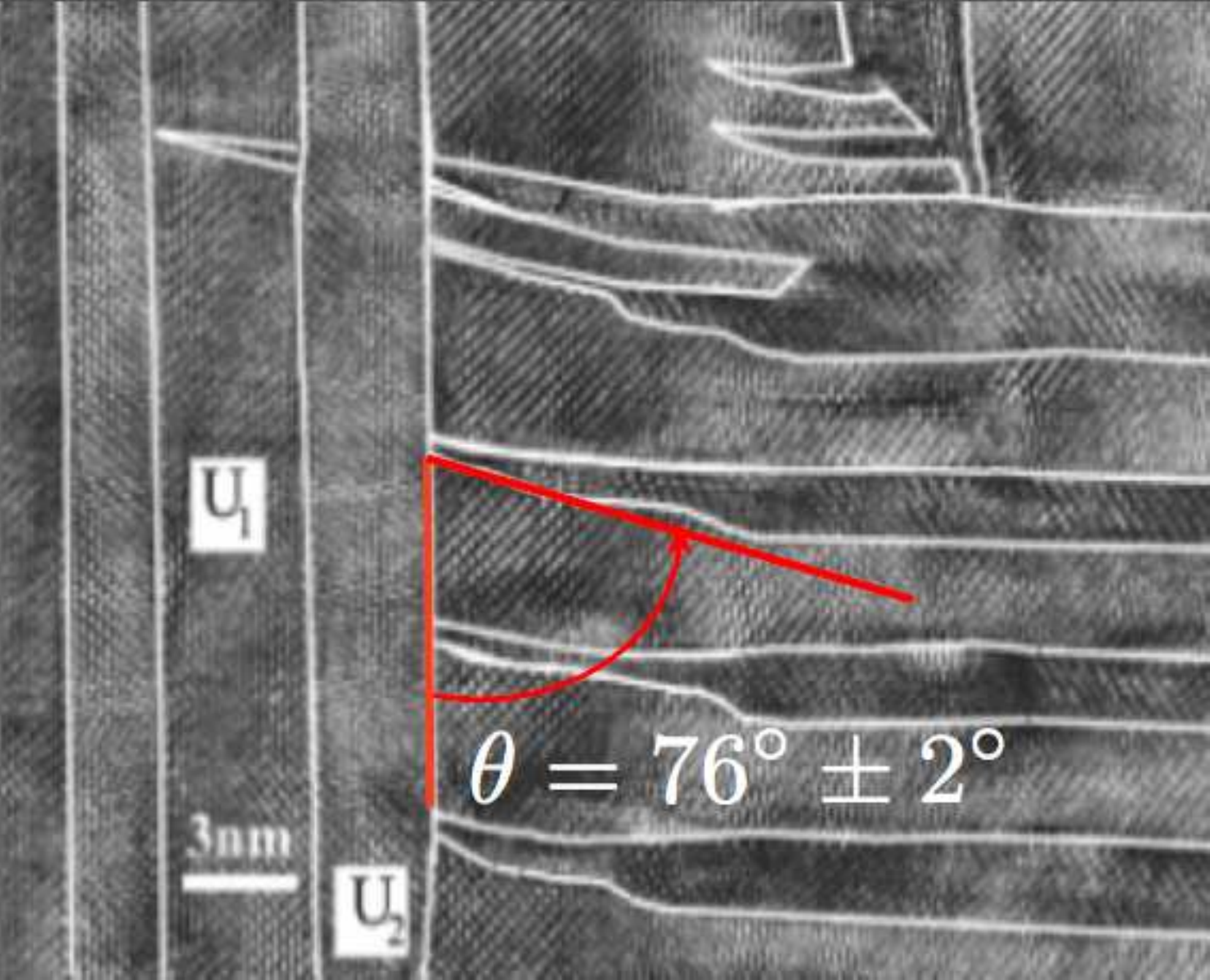}}
\vspace{-0.2cm}
\end{center}
\caption {\label{simuvsexp}Comparison between simulated microstructure and experiments. (a) Enlargement of circled region in Fig. \ref{time_evnl:edge-c}, rotated by 45\degree. (b) HRTEM image of a macrotwin boundary \cite{SchBouKohBal01}.} 
\end{figure}

In Fig. \ref{simuvsexp}, enlargement of the circled region on Fig. \ref{time_evnl:edge-c} is shown together with a HRTEM image of a macrotwin boundary in NiAl \cite{SchBouKohBal01} for a convincing comparison between the nonlinear modeling and the experimental observation. We indeed observe that the tapering and bending of alternating variants is well reproduced by the nonlinear model. We note also that needle splitting and hook-type forms are also reproduced.

%\begin{figure}
%\begin{center}
%\subfigure[][]{\label{simuvsexp:edge-a}\includegraphics[height= 4.5cm]{zoom_conf_nonlin_final.png}}
%\hspace{0.5cm}\subfigure[][]{\includegraphics[height= 4.5cm]{SchBouKohBal01B.pdf}}
%\end{center}
%\caption{\label{simuvsexp}Comparison between simulated microstructure and experiments. (a) Enlargement of circled region in Fig. \ref{time_evnl:edge-c}, rotated by 45\degree. (b) HRTEM image of a macrotwin boundary \cite{SchBouKohBal01}.} 
%\end{figure}

\section{\label{sec:level2}Discussion}

\begin{figure}
\hspace{3cm}
\begin{minipage}[t]{1\linewidth}
\includegraphics[scale=.20]{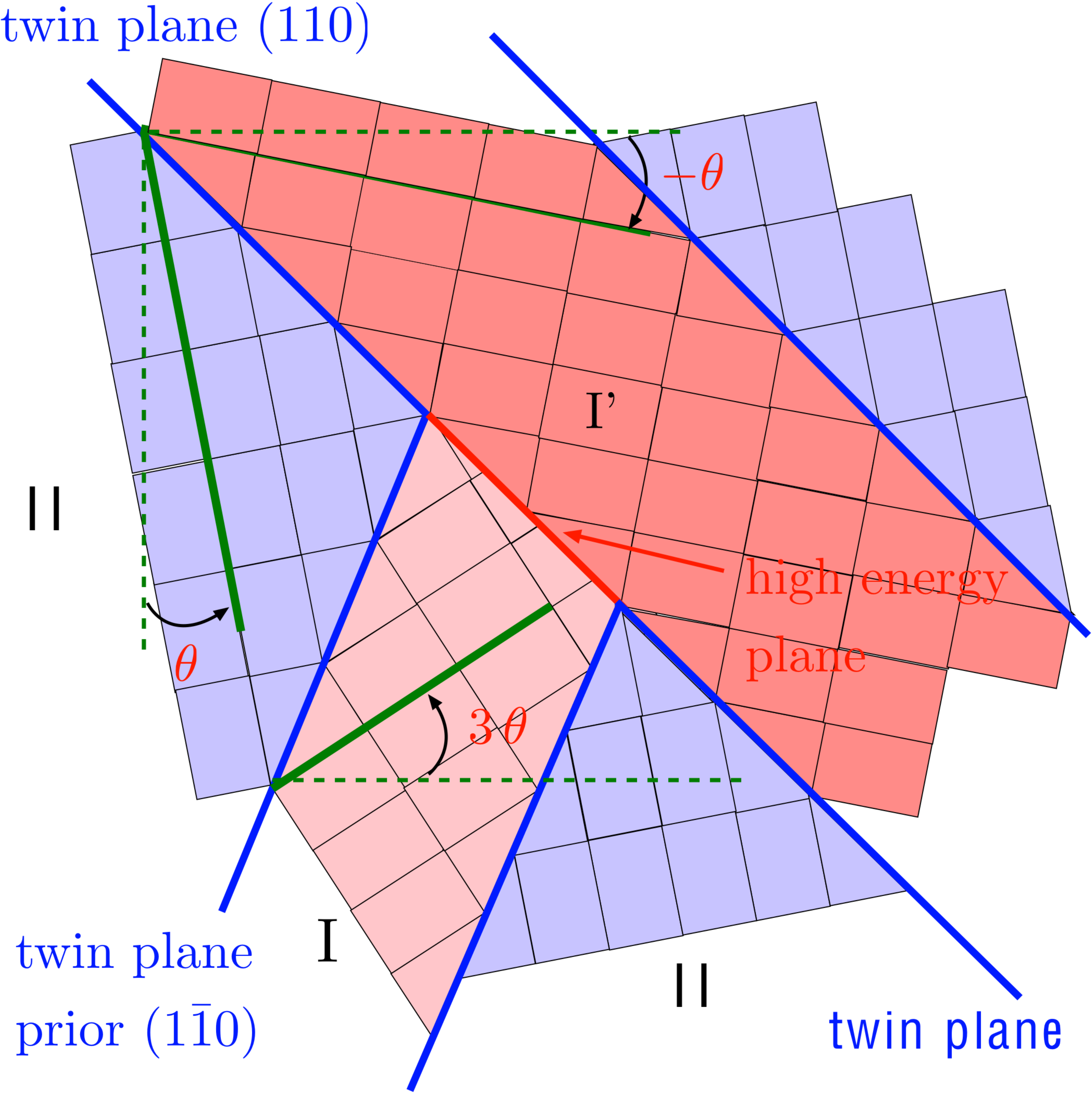}
\end{minipage}
\caption{\label{macrotwin_sketch} Schematic representation of a macrotwin boundary between two differently oriented laminates.} 
\end{figure}

\begin{figure}
\begin{center}
\subfigure[$\bold{t=30\times10^4}$][]{\label{strain_energy:edge-a}\includegraphics[height=4.5cm]{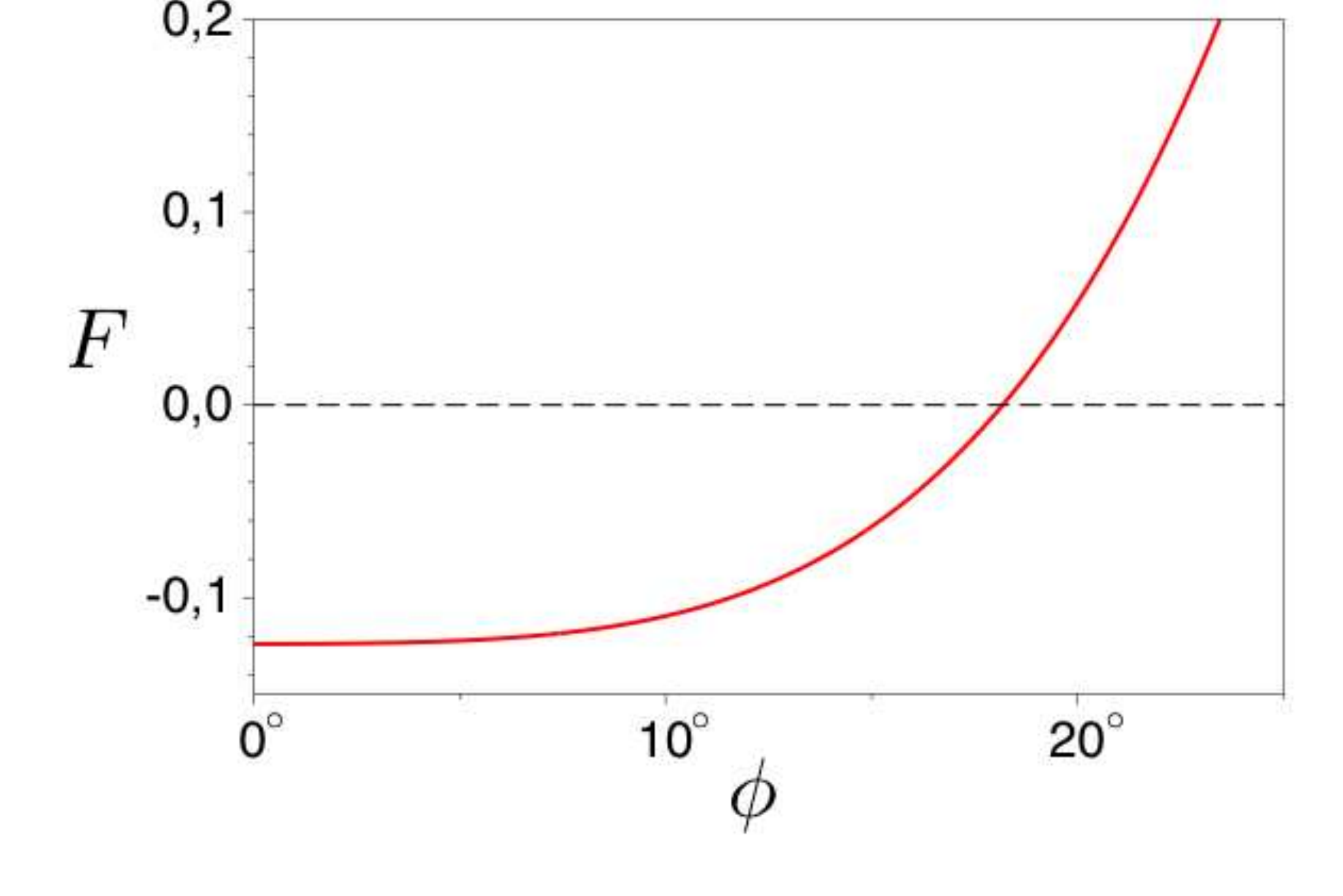}}\hspace{10mm}
\subfigure[$\bold{t=30\times10^4}$][]{\label{rotations:edge-b}\includegraphics[height= 4.5cm]{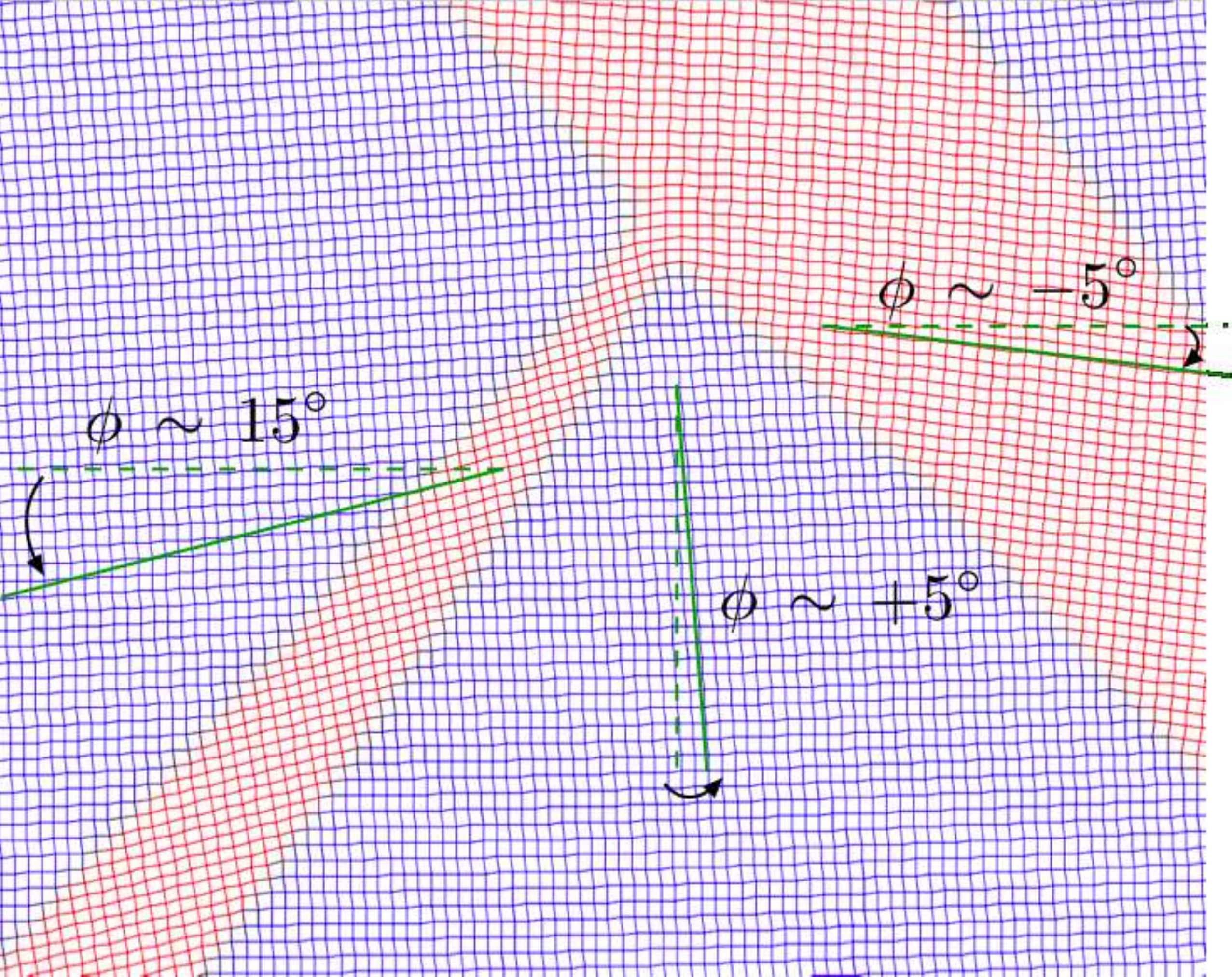}}
\caption{\label{strain_energy_function_of_phi}(a) Strain energy of an homogeneous variant as a function of the angle $\phi$ of a rigid body rotation in the linear model. (b) Simulated microstructure at grid scale in the nonlinear model. This snapshot is the magnification of the macrotwin boundary shown in Fig. \ref{simuvsexp:edge-a}. The angle $\phi$ measures the rotation of variants with respect to the reference state.  }\end{center}
\end{figure}

The driving force for the development of these microstructures is of course the reduction of the strain energy. The underlying mechanism is the following. Consider a macrotwin interface parallel to a (110) plane, as the one sketched in Fig. \ref{macrotwin_sketch}. This situation corresponds to the macrotwin boundary in the upper right corner of the final configuration displayed in Fig. \ref{time_evnl:edge-c} and also in Fig. \ref{simuvsexp:edge-a}. Accommodation, along the (110) twin plane, of variant I (on the right side of the plane) and of variant II (on the left side) require rotations equal to $-\theta$ and $+\theta$, respectively. Now, as variant II has already been rotated by $+\theta$, accommodation along the former $(1 \bar 1 0)$ twin plane that collides with the macrotwin boundary requires that variant I rotates by $3\theta$ in order to fit coherently with variant II. As a result, the former $(1 \bar 1 0)$ twin wall must rotate by $2\theta$.  This is at the origin of the bending of the twin walls that we observe in the simulation as well as in the experiment. We note also that the different rotations displayed by the variants as well as the amplitude of the bending observed in the simulation are numerically close to the expected values (see Figs. \ref{simuvsexp} and \ref{rotations:edge-b}). Another important feature of the previous accommodation mechanism is that it leads, along the macrotwin boundary, to interfaces between domains made of the same orientational variant (i.e. domains with identical deviatoric strain $e_2$) but with very different rotations. Specifically, in the situation depicted in Fig. \ref{macrotwin_sketch}, there is an interface between two domains of variant I that have been rotated by $-\theta$ and $3\theta$, respectively, leading to a high strain energy density along this interface. The system will try to shorten this high energy interface and, consequently, to lengthen the neighboring low energy twin walls. This mechanism is at the origin of tapering of alternate variants observed in the simulation and in the experiment shown in Fig. \ref{simuvsexp}.

In brief, the bending and narrowing of alternate variants along a macrotwin boundary requires large local rigid body rotations (as high as $3\theta$, with $\theta \sim 5\degree$). This illustrates the importance of using a rotationally invariant modeling, i.e. a geometrically nonlinear scheme.

Indeed, as shown in Fig. \ref{time_evl}, where we report the microstructural evolution within  the linear geometry, the time evolution with linear model is completely different. As in the nonlinear modeling, stress accommodation operates first at short length scales, as can be seen in Fig. \ref{time_evl:edge-a} where we observe the formation of needles in the neighborhood of the macrotwin boundaries. However, in further time steps,  the needles retract and martensite laminates start to disappear in favor of large martensitic domains (Fig. \ref{time_evl:edge-b}). Finally, the geometrically linear model produces a final state that extends across the whole domain, i.e., a simple laminate, shown in Fig. \ref{time_evl:edge-c}. 

The differences between the microstructures obtained within the two models can be directly linked to the role of rotations involved by the accommodation mechanism. For a quantitative understanding of this point, we present in Fig. \ref{strain_energy_function_of_phi} the variation of the strain energy density of an homogeneous variant as function of the angle $\phi$ of an applied rigid body rotation, according to the linear model, i.e. using Eqs. \ref{order_parameters} and \ref{strain_energy} with the linear relation of Eq. \ref{lin}. We remind that this strain energy density does not depend on $\phi $ if a nonlinear scheme is used. The situation is of course very different within the linear model. The strain components $\epsilon_{ij}$, and therefore the order parameters $e_i$, are no longer rotationally invariant. In particular,  a rotation of angle $\phi$ will generate non-zero dilatational and shear strains $e_1$ and $e_3$. As seen in Fig. \ref{strain_energy_function_of_phi}, the strain energy density difference between the martensite and the austenite, changes very slowly with $\phi $ for small angles, say up to approximately 10\degree. However, most of the stabilization of the martensite is lost for higher rotations. Indeed, rotations of the order of $3\theta$, as those involved in the stabilization of the needles within the nonlinear model, almost exhaust the energy difference between martensite and austenite. This unphysical behavior  is the reason why the dynamics within the linear model cannot sustain the bending and tapering of the microtwins and, therefore, cannot stabilize the macrotwin boundaries observed in the nonlinear model and in the experiment discussed above.

\begin{figure}
\begin{center}
\includegraphics[height= 4.5cm]{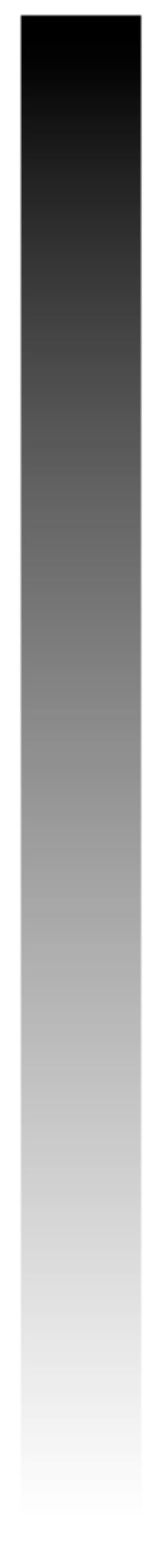}\hspace{10mm}
\subfigure[$\bold{t=30\times10^4}$][]{\label{strain_energy_maps:edge-a}\includegraphics[height= 4.5cm]{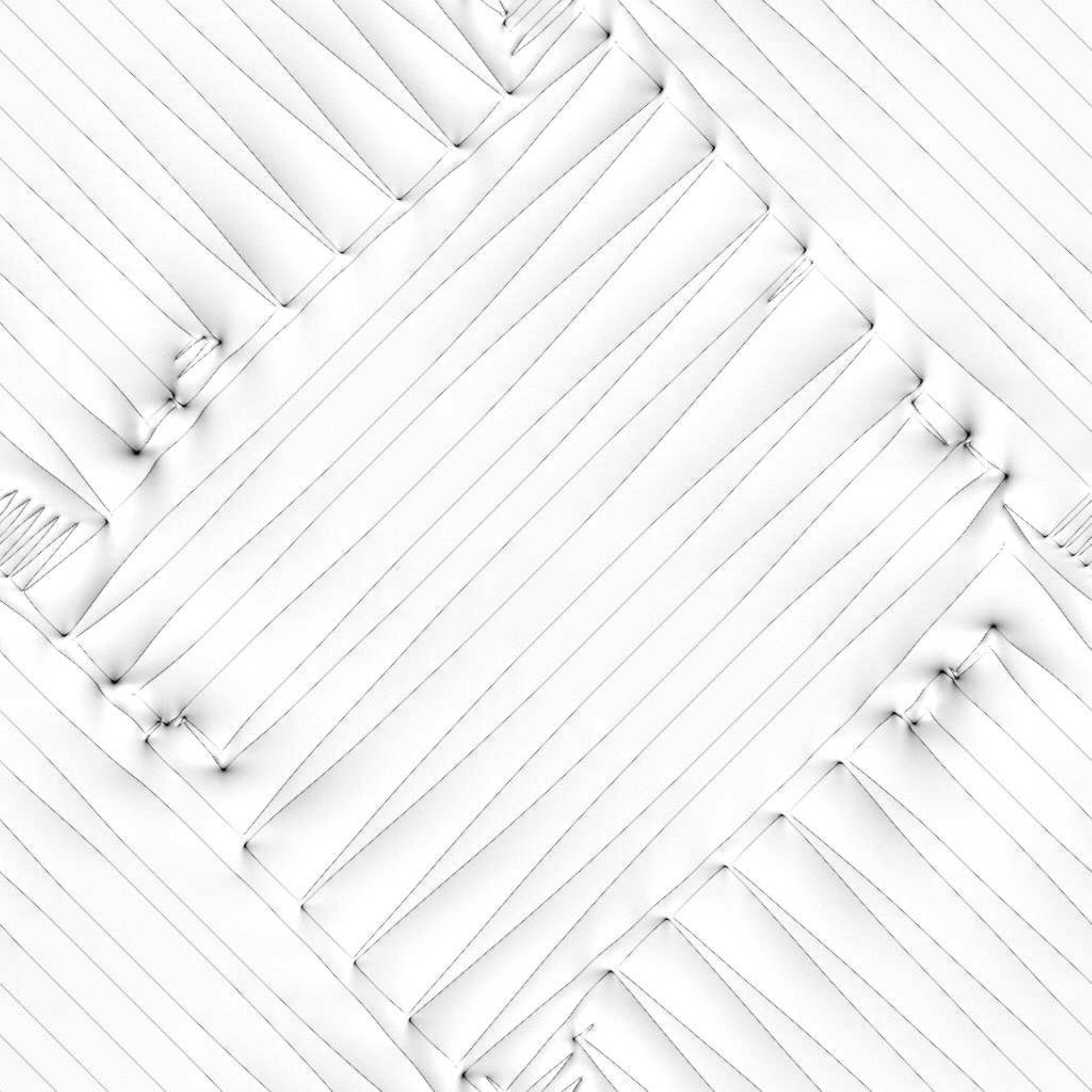}}\hspace{10mm}
\subfigure[$\bold{t=30\times10^4}$][]{\label{strain_energy_maps:edge-b}\includegraphics[height= 4.5cm]{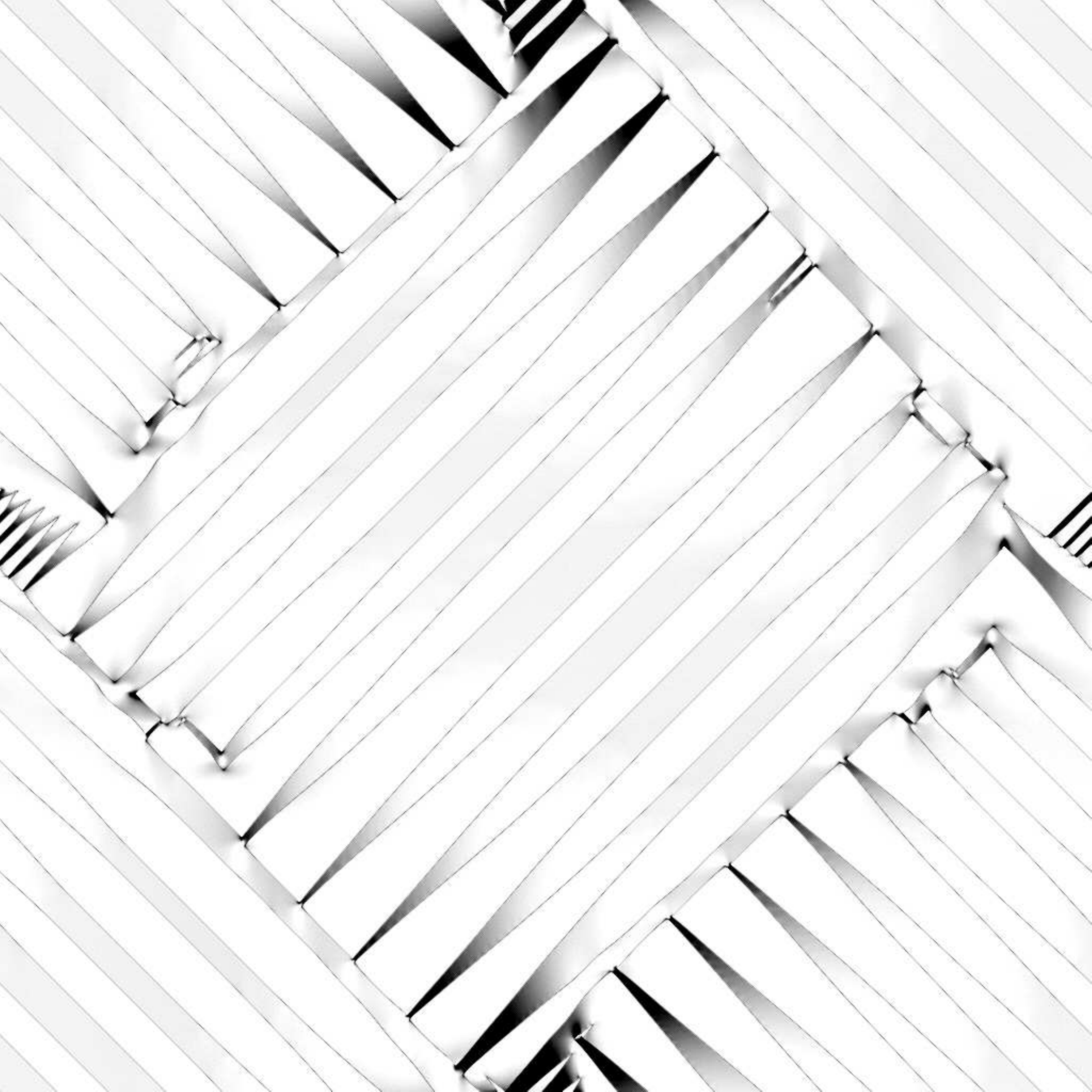}}
\caption{\label{strain_energy_maps}(a) Strain energy map of the final microstructure obtained in the nonlinear modeling. Black corresponds to the higher energy concentration, (b) strain energy map of the final microstructure obtained in the nonlinear modeling, but computed after linearization of the relation between displacement and strain.}\end{center}
\end{figure}

Another convincing illustration of this point is presented in Fig. \ref{strain_energy_maps}, where we show the strain energy density maps of the final state obtained within the exact (nonlinear) model. Two maps are presented: (a) the strain energy density  for the final (metastable) configuration obtained with the nonlinear model and, (b), the strain energy density for the same configuration, but calculated using the linear model, i.e. by replacing Eq. \ref{nonlin} by Eq. \ref{lin}. The local strain energy concentration in the exact model is locally concentrated on variant tips in very small regions (Fig.\ref{strain_energy_maps:edge-a}), precisely where the needles taper against the macrotwins, leaving short but highly energetic interfaces between identical variants. However, this high energy concentration is not strong enough to create a driving force to move the domain boundaries and consequently this metastable state can persist in the nonlinear model.  In contrast, energy concentration is high on very large regions in the linear model (Fig.\ref{strain_energy_maps:edge-b}), precisely where finite rotations are needed to accommodate variants along twin walls. As a result, the excess strain energy in the intermediate layer, where alternate variants bend, are too large for bending and splitting type microstructures to be long lived metastable states and therefore they do not persist in the linear model.

There are other computational works in the literature that claimed to observe similar microstuctures to those obtained here using the nonlinear model but none of them compared geometrically nonlinear and linear models \cite{KerEtAl99,Jac00,LevLee07} . For example, in \cite{Jac00}, a geometrically nonlinear Ginzburg-Landau approach was used to analyze twin morphologies. Using a simple conjugate-gradient minimization algorithm, twin narrowing was observed at the junction of two mutually perpendicular monodomains made of the same orientational variant. However,  large macrotwin boundaries were not analyzed and comparison with a geometrically linear model was not undertaken. Needle shape morphologies have also been observed in \cite{KerEtAl99}, using a simple geometrically linear dissipative Time Dependant Ginzburg-Landau approach. Under applied stress, nucleation of twins between previously existing variants was observed. This new twin generation exhibits needle shapes, but before they collide with the pre-existing variants. Moreover, these microstructures were only short-lived and stable macrotwin boundaries between perpendicular laminates was not observed. \cite{LevLee07} observed complicated long lived microstructures in a geometrically linear model, but no well-defined macrotwin interfaces as those analyzed here were found, and stability of multiple laminates was observed only if  a specific defect field was introduced. Concerning that point, we note that, according to the experimental observations, macrotwin interfaces do not seem to be stabilized by defects \cite{SchEtAl02}.

 \section{\label{sec:level2}Conclusion}

In summary, we have presented here a numerical comparison between two dynamical models for  understanding the formation of complex microstructures observed in materials undergoing a martensitic phase transformation. The first model  incorporates a geometrically nonlinear strain tensor, to insure that the strain energy is rotationally invariant, whereas the second one is based on its  geometrically linear approximation. The linear model is  commonly used as an approximate approach for studying the dynamics of martensitic phase transformations. Our findings show that this approximation cannot capture and reproduce the physical mechanisms at the root of the stabilization of the late stage microstructures observed in martensites. In particular, we show that the model with geometrically nonlinear strains produces final metastable states with multiple laminates, separated by stable macrotwin interfaces along which bending, tapering and splitting of microtwin needles is observed, in agreement with experiment. On the other hand, the linear model produces final states that are always simple laminates. We argue that the absence of stable macrotwin boundaries in the linear model is due to the penalty resulting from the energetically costly dilatational and shear strains generated, within a linear geometry scheme, by the lattice rotations required for the coherent accommodation of the deviatoric strain along macrotwins.

 \section*{Acknowledgements}

 We thank John Ball, Dominique Schryvers and Remi Delville for very helpful discussions.  This work was carried out with the support of the Marie Curie Research Training Network MULTIMAT (MRTN-CT-2004-505226) and the HPC-EUROPA project (RII3-CT-2003-506079).

\bibliographystyle{tPHM}
%\bibliographystyle{model3-num-names}
%\begin{thebibliography}
\bibliography{MuiSalFinBib}% Produces the bibliography via BibTeX.

\end{document}